\documentstyle[12pt,epsfig]{article}

\def\gsim{\lower0.5ex\hbox{$\:\buildrel >\over\sim\:$}}
\def\lsim{\lower0.5ex\hbox{$\:\buildrel <\over\sim\:$}}

\def \sneu{{\tilde\nu}_\mu}
\def \sneub{{\bar {{\tilde\nu}}}_\mu}

\def \sneui{{\tilde\nu}_i}

\def \sneup{{\tilde\nu}_+}
\def \sneum{{\tilde\nu}_-}
\def \sneupm{{\tilde\nu}_{\pm}}
\def \rp{{R\hspace{-0.22cm}/}_P}
\def \lp{{L\!\!\!/}}

\def \be {\begin{equation}}
\def \ee {\end{equation}}
\def \bea{\begin{eqnarray}}
\def \eea{\end{eqnarray}}

\def \n{\noindent}

\begin{document}
\baselineskip 18pt plus 2pt
\noindent \hspace*{10cm}UCRHEP-T215 (February 1998) 

\begin{center}
{\bf $R$-parity violation and CP-violating and CP-conserving 
spin asymmetries in 
\boldmath ${\ell^+\ell^- \to \tilde\nu \to \tau^+ \tau^-}$:
probing sneutrino mixing at LEP2, NLC and $\mu \mu$ colliders}
%
\vspace{.2in}

S. Bar-Shalom, G. Eilam\footnote{On leave from: Physics Department, 
Technion-Institute of Technology, Haifa 32000, 
Israel.}\\
\vspace{-5pt} 
Physics Department, University of California, Riverside, 
CA 92521, USA\\
\bigskip

and A. Soni\\
\vspace{-5pt}
Physics Department, Brookhaven National Laboratory, Upton, NY 11973,
USA\\
\vspace{.2in}

{\bf Abstract}\\
\end{center}

\n We consider the sneutrino resonance reaction 
$\ell^+ \ell^- \to \tilde\nu \to \tau^+ \tau^-$ in the MSSM without 
$R$-parity. We introduce new CP-violating 
and CP-conserving $\tau$-spin asymmetries 
which are generated already at the {\it tree-level} if 
there is $\tilde\nu - \bar {\tilde\nu}$ mixing and that are forbidden
in the SM.
It is remarkable that these spin asymmetries 
can reach $\sim 75\%$ around resonance for a sneutrino 
mass splitting of $\Delta m \sim \Gamma$  
and $\sim 10\%$ for a splitting as low as 
$\Delta m \sim 0.1 \Gamma$, where $\Gamma$ 
is the $\tilde\nu$ width. We show that they are  
easily detectable already at LEP2 so long as the beam energy  
is within $\sim 10$ GeV range around the $\sneu$ masses    
and may therefore serve
as extremely powerful probes of sneutrino mixing phenomena. 
%
%
%
\newpage 

The MSSM
lagrangian, which conserves $R$-parity, 
possesses a very distinct prediction:
superpartners must be produced in pairs and, as a 
consequence, the lightest supersymmetric particle is stable. 
This implies that direct production of sparticles is 
restricted to colliders with c.m. energy at least twice 
the typical sparticle mass. This may pose a serious 
limitation in performing a detailed study of the supersymmetric 
free-parameter space in present and future leptonic colliders. 
However, it is well known that if $R$-parity is not conserved 
(in a way that keeps the proton lifetime within its experimental
limit), then one cannot distinguish between the supermultiplets 
of the lepton-doublet 
${\hat L}$ and that of the down-Higgs doublet ${\hat H}_d$.
Thus, 
there is no good theoretical reason that prevents the superpotential from 
having additional Yukawa couplings constructed by ${\hat H}_d \to {\hat L}$ 
\cite{rpreview}. 

In this paper we are interested only in the pure 
leptonic $R$-parity violating ($\rp$) operator:  

\be
{\cal L}_{\lp} = \frac{1}{2}\lambda_{ijk} {\hat L}_i  
{\hat L}_j {\hat E}_k^c ~,
\label{rparity}     
\ee

\n that violates lepton number $L$, but not baryon number. 
${\hat E}^c$ is the charged lepton-singlet 
superfield, $i$ and $j$ are flavor indices such that 
$i \neq j$ because of the anti-commuting superfields.
${\cal L}_{\lp}$ drastically changes the 
phenomenology of the supersymmetric  
leptonic sector since it gives rise to the possibility of 
having $s$-channel slepton resonant formation in scattering 
processes, thus enabling the detection of slepton with masses 
up to the collider c.m. energy. This fact was observed already about  
10 years ago \cite{resold} and is recently gaining more interest 
for sneutrino resonance searches \cite{resnew}.

In this work we focus specifically on the effects of $\rp$ 
interactions on the process $\ell^+ \ell^- \to \tau^+ \tau^-$; 
the existence of other possible $\rp$ operators is irrelevant 
for that process at tree-level. 
In fact, the situation is rather economical for 
$e^+ e^- \to \tau^+ \tau^-$, in the sense that the fewest new 
parameters have to be considered. That is, since $i \neq j$ in 
(\ref{rparity}), the couplings 
$e {\tilde\nu}_e e$ and $\tau {\tilde\nu}_\tau \tau$ 
are absent and only the $s$-channel $\sneu$ exchange contributes 
with the couplings $\lambda_{121}$ and $\lambda_{323}$ for $e \sneu e$ 
and $\tau \sneu \tau$, respectively \cite{foot1}.
We will explore two new 
aspects of $\sneu$ resonance at LEP2: the detection of 
$\sneu - \sneub$ mixing and CP-violation. Both 
phenomena may exist once $\lambda_{121},\lambda_{323} \neq 0$ 
in (\ref{rparity}). 

Sneutrino mixing has been the subject of 
several recent papers \cite{mix1,mixyuval}.
The question of whether the sneutrinos mix or not is of 
fundamental importance since this mixing is closely 
related to the generation of neutrino masses \cite{mix1,mixyuval}. 
In fact, it was found in \cite{mixyuval} that 
$\Delta m_{\sneui} / m_{\nu_i} \lsim {\rm few} 
\times 10^3$ is required in order for $ m_{\nu_i}$ to 
be within the present experimental upper limits.   

Our present interest here is in the detection 
of sneutrino 
mixing rather than its origins. We will 
therefore not 
assume any specific model for it to occur. Instead, 
we write $\sneu = (\sneup +i \sneum)/\sqrt 2$ and 
simply assume that, due to some new short distance physics, 
there exists a mass splitting between 
the new CP-even and CP-odd muon-sneutrino mass eigenstates 
$\sneup$ and $\sneum$, respectively 
(we assume CP-conservation in the mixing), 
such that $\Delta m/m_{\pm} << 1$, 
where $\Delta m \equiv m_{+} - m_{-}$ and $m_{\pm} \equiv m_{\sneupm}$. 
In particular, we take $\Delta m \lsim \Gamma$ 
and $\Gamma \equiv \Gamma_- \simeq \Gamma_+ = 10^{-2} m_{-}$. Indeed, if 
$m_{\pm} > m_{\tilde {\chi}^+},~m_{\tilde {\chi}^0}$ 
($\tilde{\chi}^+$ and $\tilde{\chi}^0$ are the charginos and neutralinos, 
respectively), then the two-body decays 
$\sneupm \to \tilde{\chi}^+ \ell;~\tilde{\chi}^0 \nu$ are open and the 
corresponding partial widths are given by:  
$\Gamma(\sneupm \to \tilde{\chi}^+ \ell);~\Gamma(\sneupm \to 
\tilde{\chi}^0 \nu) 
\sim {\cal O} \left[ 10^{-2} m_{\pm} \times 
\left(1 - m_{\tilde {\chi}^+}^2/m_{\pm}^2 \right)^2;~\left(1-
m_{\tilde {\chi}^0}^2/m_{\pm}^2\right)^2 \right]$ 
(see Barger {\it et al.} in \cite{resold}). 
Therefore, for $m_{\pm} \gsim 200$ GeV,  
$\Gamma= 10^{-2} m_{-}$ serves our purpose as it 
is a viable estimate
even without taking into account the new $\rp$ 
two-body decay modes which, when summed, can also form a significant 
fraction of the total sneutrino width. 

Apart from the rough theoretical requirement 
that $\Delta m/m_{-} << 1$, imposed by neutrino masses 
\cite{mixyuval}, there is another important 
reason why we wish to consider the limit $\Delta m \lsim \Gamma$. 
The reason is that in that 
case the $\sneup$ and $\sneum$ resonances will overlap and 
distinguishing between the two peaks becomes a 
non-trivial experimental task. 
In such an event, in order to observe 
the small $\sneup-\sneum$ mass splitting 
one would have to search for the effects of flavor oscillations
in sneutrino decays in analogy to the $B^0-\bar{B}^0$ system, for example   
in $\tilde\nu$-pair production, $e^+e^- \to \sneup \sneum$ \cite{mixyuval}.   
However, for $m_{\pm} \gsim 200$ GeV, 
such a $\tilde\nu$-pair production awaits the next generation 
of lepton colliders.
In what follows, we will show that an alternative for a 
detection of $\Delta m \neq 0$ for $m_{\pm} \gsim 200$ GeV
 is to measure appropriate CP-even and CP-odd $\tau$-spin 
asymmetries in $e^+e^- \to \tau^+ \tau^-$ which are proportional to 
$\Delta m$. We show that these asymmetries may reach tens of 
percents in a relatively wide sneutrino mass range around the sneutrino 
resonance even for a small splitting of $\Delta m \lsim \Gamma /4$. 
As a consequence, such a small mass splitting may be detectable already 
at LEP2 to many standard deviations (SD). Note that spin asymmetries in 
$\ell^+ \ell^- \to f \bar f$ can be measured only for 
$f=\tau~{\rm or}~t$. However, in sneutrino resonant formation they may 
apply only to the $\tau$ since the $t \tilde\nu t$ coupling is 
forbidden by gauge invariance. 
 
Besides, the possibility of having tree-level CP-violation,
of
the 
order of tens of percent in $\tau$ pair production  
at LEP2, 
stands out as an extremely interesting issue by itself. 
Previous studies of CP-violating effects in $e^+e^- \to \tau^+ \tau^-$,  
that can emanate 
from models beyond the SM, 
such as 
multi-Higgs 
doublet model, SUSY, 
leptoquark and Majorana $\tilde\nu$, all involve one-loop exchanges of 
the new particles which 
generate 
a CP-violating electric dipole moment 
for the $\tau$ (see \cite{cptau} and references therein). 
These CP-odd effects
are therefore much smaller than our tree-level effect, 
which may exist at a level of tens of percents around the 
$\sneupm$ resonance.    

Let us now construct the $\tau^+ \tau^-$ double-polarization asymmetries. 
In the rest frame of $\tau^-$ we define the basis vectors: 
$\vec{e}_z \propto - ({\vec{p}}_{e^+} + {\vec{p}}_{e^-})$, $\vec{e}_y \propto 
{\vec{p}}_{e^+} \times {\vec{p}}_{e^-}$ and     
$\vec{e}_x = \vec{e}_y \times \vec{e}_z$. 
For the $\tau^+$ we use a similar set of definitions such that 
$\vec{{\bar e}}_x, \vec{{\bar e}}_y, \vec{{\bar e}}_z$ 
are related to $\vec{e}_x,\vec{e}_y,\vec{e}_z$ 
by charge conjugation.
 We then define the following $\tau^+\tau^-$ double-polarization 
operator with respect to their corresponding rest frames defined above:

\bea
\Pi_{ij} \equiv \frac{ N({\bar {\uparrow}}_i \uparrow_j) - 
N({\bar {\uparrow}}_i \downarrow_j) - N({\bar {\downarrow}}_i \uparrow_j) +
N({\bar {\downarrow}}_i \downarrow_j) }{ N({\bar {\uparrow}}_i \uparrow_j) + 
N({\bar {\uparrow}}_i \downarrow_j) + N({\bar {\downarrow}}_i \uparrow_j) +
N({\bar {\downarrow}}_i \downarrow_j)} ~, \label{piij} 
\eea  

\n where $i,j=x,y,z$. For example, 
$N({\bar {\uparrow}}_x \uparrow_y)$ stands 
for the number of events in which $\tau^+$ has 
spin +1 in the direction $x$ in its rest frame 
and $\tau^-$ has spin +1 in the direction $y$ in its 
rest frame. The spin vectors of $\tau^+$ and $\tau^-$ 
are therefore defined in their respective rest frames as: 
${\vec s}\,^+= ({\bar s}_x,{\bar s}_y,{\bar s}_z)$ and ${\vec s}\,^-= 
(s_x,s_y,s_z)$ and  
$\Pi_{ij}$ is calculated in the $e^+e^-$ c.m. frame by boosting 
${\vec s}\,^+$ and ${\vec s}\,^-$ from the $\tau^+$ and $\tau^-$ rest 
frames to the $e^+e^-$ c.m. frame.

It is then easy to verify that $\Pi_{ij}$ possesses the 
following transformation properties under the operation 
of CP and 
of
the naive time reversal $T_N$ \cite{foot2}:
 $CP(\Pi_{ij})=\Pi_{ji}$ 
for all $i,j$, 
$T_N(\Pi_{ij})=-\Pi_{ij}$ for $i~{\rm or}~j=y$ and $i \neq j$ and      
$T_N(\Pi_{ij})=\Pi_{ij}$ for $i,j \neq y$ and for $i=j$. 
We can therefore define:

\bea
A_{ij}=\frac{1}{2} \left(\Pi_{ij}-\Pi_{ji} \right) ~~,~~
B_{ij}=\frac{1}{2} \left(\Pi_{ij}+\Pi_{ji} \right)~.\label{asym}
\eea

\n Evidently, $A_{ij}$ are CP-odd ($A_{ii}=0$ by definition) 
and $B_{ij}$ are CP-even. Also, $A_{xy},A_{zy},B_{xy}$ and
$B_{zy}$ are 
$T_N$-odd while $A_{xz},B_{xz},B_{xx},B_{yy}$ and $B_{zz}$ are $T_N$-even.  

To calculate the various asymmetries defined in (\ref{asym}) we need 
the cross-sections for the $s$-channel sneutrino exchange and the SM $s$-channel 
$\gamma,Z$ exchanges. The interferences between the SM diagrams 
and the $s$-channel $\sneupm$ diagrams as well as between the $\sneup$ and 
the $\sneum$ $s$-channel diagrams are proportional to the electron mass 
and are therefore being neglected. The SM and $\sneupm$ cross sections can be 
subdivided as: $\sigma_{SM;\sneupm} \equiv \sigma_{SM;\sneupm}^0/4 + 
\sigma^{{\vec s}\,^-{\vec s}\,^+}_{SM;\sneupm}$, where $\sigma_{SM}^0$ 
and $\sigma_{\sneupm}^0$ are the SM and $\sneupm$ total cross sections,  
respectively, summed over the $\tau^+$ and $\tau^-$ spins; 
$\sigma^{{\vec s}\,^-{\vec s}\,^+}_{SM;\sneupm}$ are the spin 
dependent parts.  
The total spin dependent cross-section for $e^+ e^- \to \tau^+ \tau^-$ 
is then simply given by the sum 
$\sigma^T=\sigma_{SM}+\sigma_{\sneupm}$.
For the SM we find (assuming $m_{\tau}=0$ and $\Gamma_Z/m_Z=0$):

\bea
\sigma_{SM}^0&=&\frac{\pi \alpha^2}{3s} \left(4+2\omega 
(g_L+g_R)^2 +\omega^2(g_L^2+g_R^2)^2 \right)~,\\
\sigma^{{\vec s}\,^-{\vec s}\,^+}_{SM}&=&\frac{\pi \alpha^2}{12s} 
\left\{ s_z {\bar s}_z \left(4+2\omega (g_L+g_R)^2 +
\omega^2(g_L^2+g_R^2)^2 \right) + \right. \nonumber\\ 
&&\left. (s_x {\bar s}_x - s_y {\bar s}_y) \left(2+
\omega (g_L+g_R)^2 +\omega^2 g_Lg_R(g_L^2+g_R^2)\right) + 
\right. \nonumber\\
&&\left. (s_z + {\bar s}_z) \left(2\omega (g_R^2-g_L^2) +
\omega^2 (g_R^4-g_L^4)\right) \right\}~, \label{sm}
\eea

\n where $s=(p_{e^+} + p_{e^-})^2$, $\omega \equiv 
\left(\sin^2 \theta_W \cos^2 \theta_W (1-m_Z^2/s) \right)^{-1}$, 
$g_L=\sin^2 \theta_W -1/2$, $g_R=\sin^2 \theta_W$  
and $\theta_W$ is the weak mixing angle. 

For the $s$-channel $\sneupm$ we assume for simplicity that 
$\lambda_{121}$ is real (this assumption does not change our 
predictions below) and define 
$\lambda_{323} \equiv (a+ib)/\sqrt 2$. The relevant couplings 
of the CP-even ($\sneup$) and the CP-odd ($\sneum$) sneutrino 
mass eigenstates are then: $e \sneup e = i \lambda_{121}/\sqrt 2$, 
$e \sneum e = - \lambda_{121} \gamma_5/\sqrt 2$, 
$\tau \sneup \tau = i (a-ib \gamma_5)/2$, $\tau \sneum \tau = 
i (b+ia\gamma_5)/2$ 
and the sneutrinos cross-section is ($m_\tau=0$):

\bea
\sigma_{\sneupm}^0&=&\frac{s}{64 \pi} \lambda_{121}^2 
|\lambda_{323}|^2 D_+ ~,\\ 
\sigma^{{\vec s}\,^-{\vec s}\,^+}_{\sneupm}&=& - 
\frac{s}{512 \pi} \lambda_{121}^2 
\left\{ s_z {\bar s}_z (a^2+b^2) D_+ + 
(s_x {\bar s}_x + s_y {\bar s}_y) (b^2-a^2) D_- + 
\right. \nonumber \\       
&& \left. 2ab (s_y {\bar s}_x - s_x {\bar s}_y) D_- 
\right\}~, \label{sneu}
\eea 

\n where $D_\pm \equiv |\pi_+|^2 \pm |\pi_-|^2$ and $\pi_\pm = 
\left(s -m_{\pm}^2 +i m_{\pm} \Gamma \right)^{-1}$. 

The calculation of the various spin asymmetries is now straightforward. 
For the $\sneupm$ exchanges, at tree-level, 
 only $A_{xy},B_{xx},B_{yy}$ and $B_{zz}$ are non-zero:

\bea
{\rm Sneutrinos~only:}~~A_{xy}=\left(\frac{2ab}{a^2+b^2}\right) 
\frac{D_-}{D_+}~~,~~B_{xx}=B_{yy}=
\left(\frac{a^2-b^2}{a^2+b^2}\right) \frac{D_-}{D_+}~~,~~B_{zz}=-1~.\label{sneuasym}
\eea

\n For the SM case, only the following CP-even 
asymmetries are non-zero at tree-level:

\bea  
{\rm SM~only:}~~B_{xx}=-B_{yy}=\frac{2+
\omega (g_L+g_R)^2 +\omega^2 g_Lg_R(g_L^2+g_R^2)}
{4+2\omega (g_L+g_R)^2 +\omega^2 (g_L^2+g_R^2)^2}~~,~~B_{zz}=1~.\label{smasym}
\eea

As expected, a non-vanishing CP-odd asymmetry is unique to the sneutrino 
exchange contribution. It is clear from (\ref{sneu}) and (\ref{sneuasym}) 
that CP-violation in $e^+e^- \to \sneupm \to \tau^+ \tau^-$ 
arises already at tree-level from the interference of the 
scalar and pseudoscalar couplings 
of $\sneupm$ to $\tau^+\tau^-$. The possibility of 
generating a tree-level CP-violating effect when 
a scalar-fermion-antifermion coupling is of the form 
$(a+ib \gamma_5)$ was first 
observed in \cite{ourtth}. There it was suggested that 
a neutral Higgs of a two Higgs doublet model may drive 
such large tree-level CP-violating effects.       
Recently, it was shown \cite{bern} that spin correlations can trace 
similar scalar-pseudoscalar tree-level interference effects in
the $H^0 \to t \bar t$ and $H^0 \to \tau^+ \tau^-$ decay modes.
However, the tree-level CP-violation effect in 
$H^0 \to \tau^+ \tau^-$, when applied to $e^+e^- \to H^0 \to \tau^+\tau^-$, 
is only of academic interest since the neutral 
Higgs coupling to electrons is $\propto m_e$.

In \cite{bern}, a non-vanishing tree-level 
CP-even spin correlation of the form ${\vec s}\,^+ \cdot {\vec s}\,^-$ was 
suggested for $H^0 \to \tau^+ \tau^-$. However, we note that  
${\vec s}\,^+ \cdot {\vec s}\,^-$ simply translates to the observable 
$\Sigma_{i=x,y,z} B_{ii}$ and it is therefore 
clear from (\ref{smasym}) that in the SM,  
${\vec s}\,^+ \cdot {\vec s}\,^- \propto B_{zz} =1$. A measurement 
of such an observable will therefore be insensitive to the couplings 
$a$ and $b$ in $\lambda_{323}$.
We suggest here a way out by defining the new CP-even 
observable: $B \equiv (B_{xx}+B_{yy})/2$. Obviously, at tree-level, 
$B=0$ in the SM and $B=B_{xx}=B_{yy}$ for the sneutrino case. 
Thus, a measurement of $B \neq 0$ will be a strong 
indication for the existence 
of new physics in 
$e^+e^- \to \tau^+ \tau^-$ in the form of  
new non-vanishing $s$-channel scalar exchanges and        
will provide explicit information 
on the new $\tau \sneu \tau$ coupling.

 From (\ref{sneuasym}) we 
observe that $A_{xy}~{\rm and}~B \propto D_-/D_+$ where 
the proportionality 
factors do not depend on the absolute magnitude of the couplings 
$a$ and $b$ but rather on any function of their 
ratio $f(a/b)$. 
In particular, without loss of generality, we will assume that 
$a$ and $b$ are positive and study the 
asymmetries as a function of the ratio $r \equiv b/(a+b)$. 
$r$ can vary between $0 \leq r \leq 1$, 
where the lower and upper limits of $r$ are given by $b=0$ and 
$a=0$, respectively. One can immediately observe that $A_{xy}$ and $B$ 
complement each other as they probe opposite ranges of $r$.
For $A_{xy}$ the maximal value $D_-/D_+$
is obtained when 
$r=1/2$ ($a=b$) and $B=D_-/D_+$ when $r=0$ ($b=0$). Also, at 
$r=1$ ($a=0$), $B=-D_-/D_+$, thus reaching its maximum negative value.

In Figure~1 we plot the ratio $D_-/D_+$, 
{\it i.e.}, the maximal values of $A_{xy}$ and $B$, as a function 
of the lighter muon-sneutrino mass $m_-$. 
We take 
$\Delta m =\Gamma,~
\Gamma /2,~\Gamma /4,~\Gamma /10$ 
(recall that $\Gamma = 10^{-2} m_{-}$). 
Also, here and throughout the rest of the paper we take the c.m. energy 
at LEP2 to be $E_{CM}=192$ GeV. 
Evidently, $A_{xy}$ and $B$ can reach $\sim 75\%$ around 
resonance if $\Delta m =\Gamma$, and $\sim 10\%$ even for the 
very small splitting $\Delta m = \Gamma /10$. 
We also observe 
that the asymmetries stay large ($\gsim 10\%$) 
even $\sim 10$ GeV away from resonance.  
Around the narrow region of  
$E_{CM} \sim (m_{+}+m_{-})/2$, $D_-/D_+ \simeq \Delta m/m_{-}$ 
and the asymmetries become very small. 

The statistical significance, $N_{SD}$, 
with which $A_{xy}~{\rm or}~B$ can be detected, is given 
by $N_{SD} = \sqrt N |{\cal A}| \sqrt {\epsilon}$, where 
${\cal A}=A_{xy}~{\rm or}~B$, 
$N=(\sigma^0_{\sneupm}+\sigma_{SM}^0) \times L$ 
is the total number 
of $e^+e^- \to \tau^+ \tau^-$ events and we take 
$L=0.5$ fb$^{-1}$ 
as the total integrated luminosity at LEP2. 
$\epsilon$ is 
the combined 
efficiency for the simultaneous measurement of
 the $\tau^+$ and $\tau^-$ spins which, therefore,   
depends on the efficiency for the spin analysis
 and also on the branching ratios 
of the specific $\tau^+$ and $\tau^-$ decay channels that are being analyzed. 
The simplest examples perhaps are the two-body 
decays $\tau^\pm \to \pi^\pm \nu_\tau$ and $\tau^\pm \to \rho^\pm \nu_\tau$, 
although 3-body decays may also be useful 
\cite{bern,tsay}. When all combinations of only the above $\tau^+,\tau^-$ 
two-body decay channels are taken into account 
one finds $\epsilon \sim 0.03$ \cite{bern}.
We will adopt this conservative number henceforward.   

In Figure 2 we plot the statistical significance, $N_{SD}$,
which corresponds to $A_{xy}$ and $B$ 
at their maximal values, at LEP2, as a function of $m_-$. 
We choose the same values for $\Delta m$ as in Figure 1. 
For completeness, the SM contribution to the denominator 
in (\ref{piij}) is now being included, in which case $A_{xy}$ 
and $B=B_{xx}=B_{yy}$ in (\ref{sneuasym}) are 
multiplied by the factor 
$\left( 1 + \sigma^0_{SM}/\sigma^0_{\sneupm} \right)^{-1}$.  
We calculate $\sigma^0_{\sneupm}$ by setting $\lambda_{121}$ 
and $\lambda_{323}$ to their presently experimental allowed 
upper limits \cite{rpreview}: 
$\lambda_{121}=0.05 \times (m_- /100 ~{\rm GeV})$ and 
$|\lambda_{323}|=0.06 \times (m_- /100 ~{\rm GeV})$. 
It is remarkable that both $A_{xy}$ and $B$  
may be detectable, under the best circumstances,
with a sensitivity reaching well above 
$\sim 10$ SD. 
We see, for example, that for 
$\Delta m =\Gamma$ these 
asymmetries induce beyond 
a 3$\sigma$ effect at LEP2, around the resonance region, 
practically over the 
whole $\sim 10$ GeV mass range,  
$186.5~{\rm GeV} \lsim m_- \lsim 196 ~{\rm GeV}$. 
For $\Delta m =\Gamma/4$
the corresponding 3$\sigma$ mass range is 
$189.5~{\rm GeV} \lsim m_- \lsim 194 ~{\rm GeV}$ and even 
for the very small splitting $\Delta m =\Gamma/10$ 
there is a $3 \sigma$ region over about a 1 GeV 
interval near the resonance mass.

Figure 3 shows the dependence of $N_{SD}$, for $A_{xy}$ and $B$,  
on the ratio $r$, where, as in Figure 2, the SM diagrams are included
and for illustration we set $m_-=E_{CM}=192$ GeV (we note that this 
value of $m_-$ does not maximize the effects). We see that a measurement of
$A_{xy}$ and $B$ at LEP2 can cover a wide range of the parameter $r$. 
In particular, one observes that for $\Delta m= \Gamma$, LEP2 
can have larger than $3 \sigma$ sensitivity to values of $r$ 
practically over the entire range of $r$ for both $A_{xy}$ and $B$.
Even for $\Delta m= \Gamma /4$, the following ranges are 
covered to at least a 3$\sigma$ significance: for $A_{xy}$,  
 $0.27 \lsim r \lsim 0.73$ and for $B$, $0 \leq r \lsim 0.32$ 
and $0.68 \lsim r \leq 1$. Evidently, as stated before, 
the ranges being covered by $A_{xy}$ and $B$ complement each other 
such that, even for the $\Delta m= \Gamma /4$ case, the whole range 
$0 \leq r \leq 1$ can be covered, to at least 3$\sigma$, 
with the simultaneous measurement of 
$A_{xy}$ and $B$. 
We note again that the $r$ ranges being covered to at least 3$\sigma$ 
by each of the two asymmetries are wider if $m_-$ is slightly away 
from resonance, {\it i.e.}, by about $0.5$ GeV. 

Finally, we have calculated the sensitivity of the NLC with c.m. energy 
$E_{CM}=500$ GeV to $A_{xy}$ and $B$ for a very 
heavy muon-sneutrino $m_- \sim E_{CM}$. 
We found that the NLC will 
be able to probe these CP-odd and the CP-even asymmetries to 
at least 3$\sigma$ (the best effects are again at the $\sim$ 20$\sigma$ 
level),
in the muon-sneutrino mass range of $\sim 20$ GeV  around resonance, 
{\it i.e.}, 
$E_{CM} -10 ~{\rm GeV} \lsim m_- \lsim E_{CM} +10 ~{\rm GeV}$, even 
for a small mass splitting of $\Delta m = 1 ~{\rm GeV} \simeq \Gamma /5$. 
Also, we found that, with $\Delta m = 1 ~{\rm GeV}$, the NLC will have 
a sensitivity above 3$\sigma$ to either $A_{xy}$ or $B$ 
over almost the entire $r$ 
range, $0 \leq r \leq 1$.      
     
To summarize, we have introduced new 
CP-violating and CP-conserving spin asymmetries and applied them 
to $e^+ e^- \to \tau^+ \tau^-$. We have shown that two 
of these asymmetries are unique in 
their ability to distinguish between the CP-odd and CP-even muon-sneutrino 
mass eigenstates in $e^+ e^- \to \sneupm \to \tau^+ \tau^-$. Both 
asymmetries arise already at the tree-level and can become 
extremely large, 
of the order of tens of percent. They may therefore be 
detectable with many SD's 
already at LEP2 if the muon-sneutrino mass lies within $\sim 10$ GeV 
range around the 
LEP2 c.m. energy, even if the mass splitting between the two 
sneutrino particles is less than $1$ GeV. 
As far as CP-violation is 
concerned, it is especially 
gratifying that such a large CP-nonconserving effect may arise in 
$\tau$-pair production 
at LEP2 
and may be searched for in the very near future.

We have also found that these asymmetries will yield a 
significant 
signal 
at the NLC with $E_{CM}=500$ GeV within a wider sneutrino 
mass range, of $\sim 20$ GeV, around resonance.
Moreover, 
the effects reported here may be similarly applied to a future 
muon collider in the ${\tilde\nu}_e$ resonance channel 
$\mu^+ \mu^- \to {\tilde\nu}_e \to \tau^+ \tau^-$. However, 
we note that while the present limits on the 
$\tau {\tilde\nu}_e \tau$ and $\tau {\tilde\nu}_\mu \tau$ couplings 
are comparable, the limit on the coupling 
$\mu {\tilde\nu}_e \mu$ is more stringent then the one on 
$e {\tilde\nu}_\mu e$ by about an order 
of magnitude.                 
    
In parting we wish to remark that  
a measurement of the double $\tau$-spin 
asymmetries $A_{xy}$ and $B$ in $\tau^+ \tau^-$ production 
at the Tevatron is another interesting possibility 
\cite{tevatron}.

\bigskip\bigskip

We acknowledge partial support from U.S. Israel BSF (G.E. and A.S.) 
and from the U.S. DOE contract numbers DE-AC02-76CH00016(BNL), 
DE-FG03-94ER40837(UCR). S.B. thanks J. Wudka and D.P. Roy for 
helpful discussions. G.E. thanks the Israel Science Foundation and the Fund 
for the Promotion of Research at the Technion for partial support 
and members of the HEP group in UCR for their hospitality. 
\pagebreak

\pagebreak

\begin{center}
{\bf Figure Captions}
\end{center}

\begin{description}

\item{Fig. 1:} The maximal value of $A_{xy}$ and $B$, {\it i.e.} 
$D_-/D_+$,   
as a function of the lighter $\sneu$ mass $m_-$, for 
four mass-splitting values $\Delta m$.

\item{Fig. 2:} The statistical significance, $N_{SD}$, 
attainable at LEP2 for $A_{xy}$ and $B$ at their maximal values, 
as a function of $m_-$. See also caption to Figure 1. 
  
\item{Fig. 3:} The attainable $N_{SD}$,  
for $A_{xy}$ and $B$, at LEP2, 
as a function of $r \equiv b/(a+b)$. The cases $\Delta m= \Gamma$ and 
$\Delta=\Gamma/4$ are illustrated.
See also caption to Figure 1. 
  
\end{description}

\newpage
\pagestyle{empty}

\begin{figure}
\centering
\leavevmode
\epsfysize=500pt
\epsfbox[0 0 612 792]{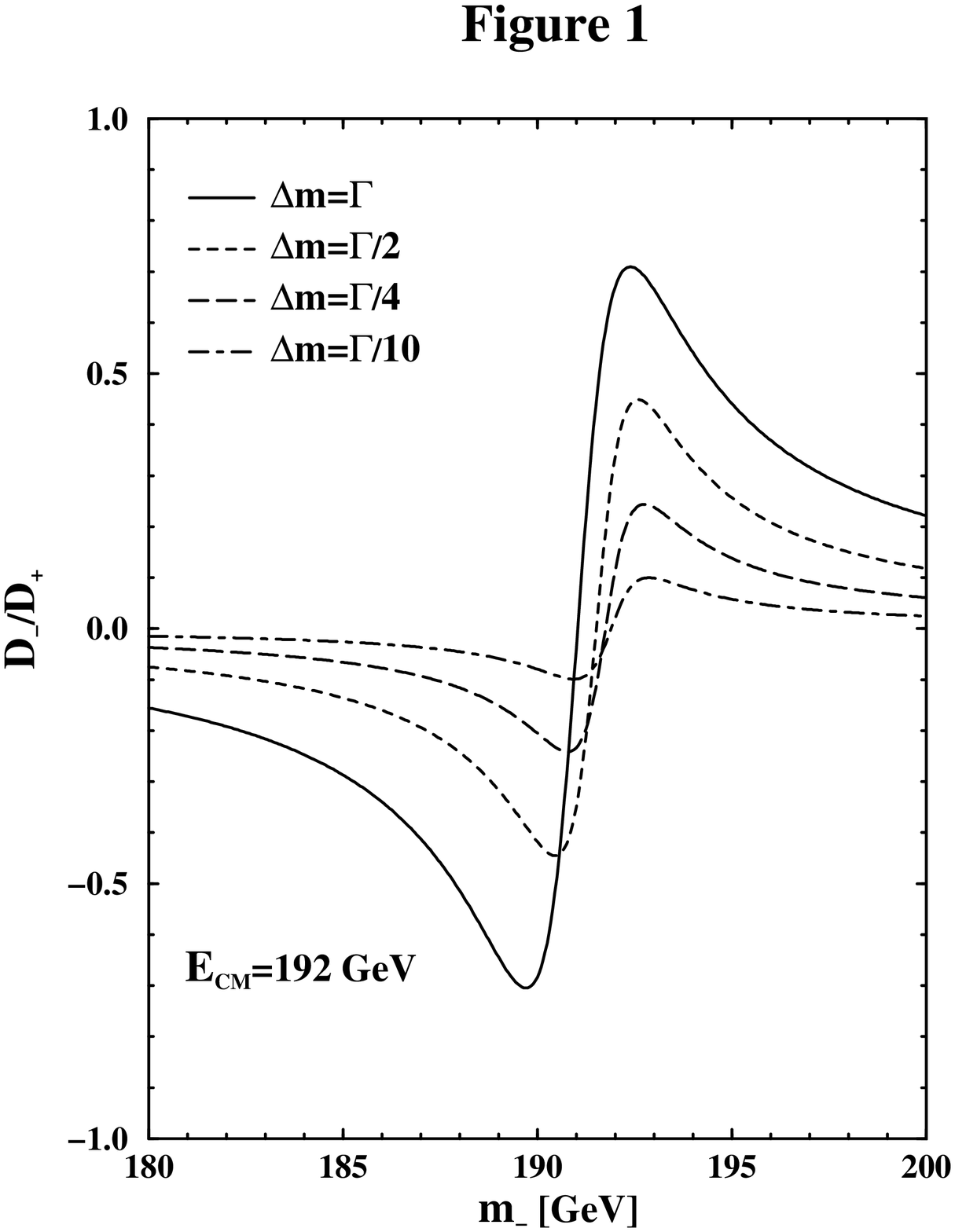}
\end{figure}

\begin{figure}
\centering
\leavevmode
\epsfysize=500pt
\epsfbox[0 0 612 792]{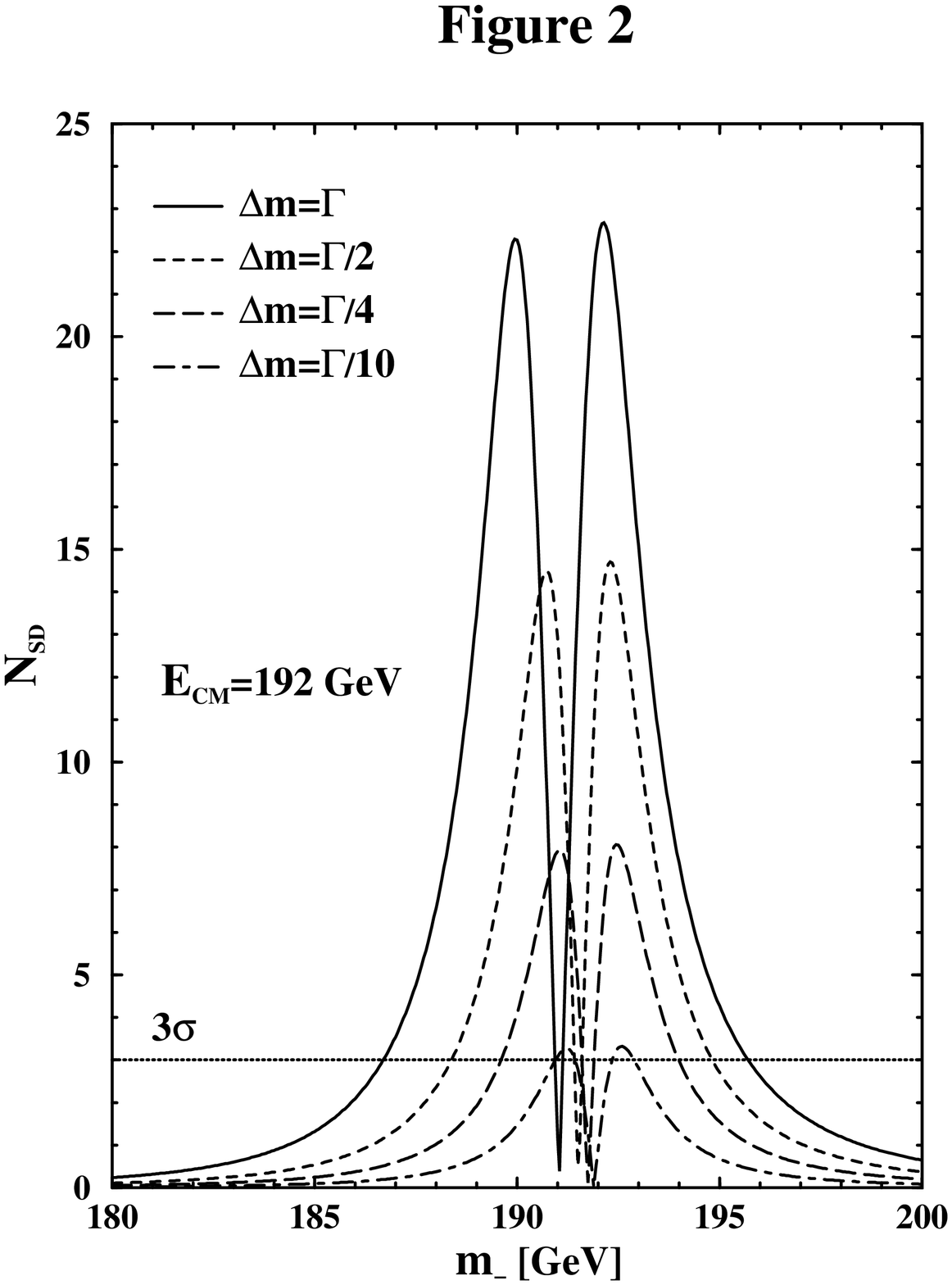}
\end{figure}

\begin{figure}
\centering
\leavevmode
\epsfysize=500pt
\epsfbox[0 0 612 792]{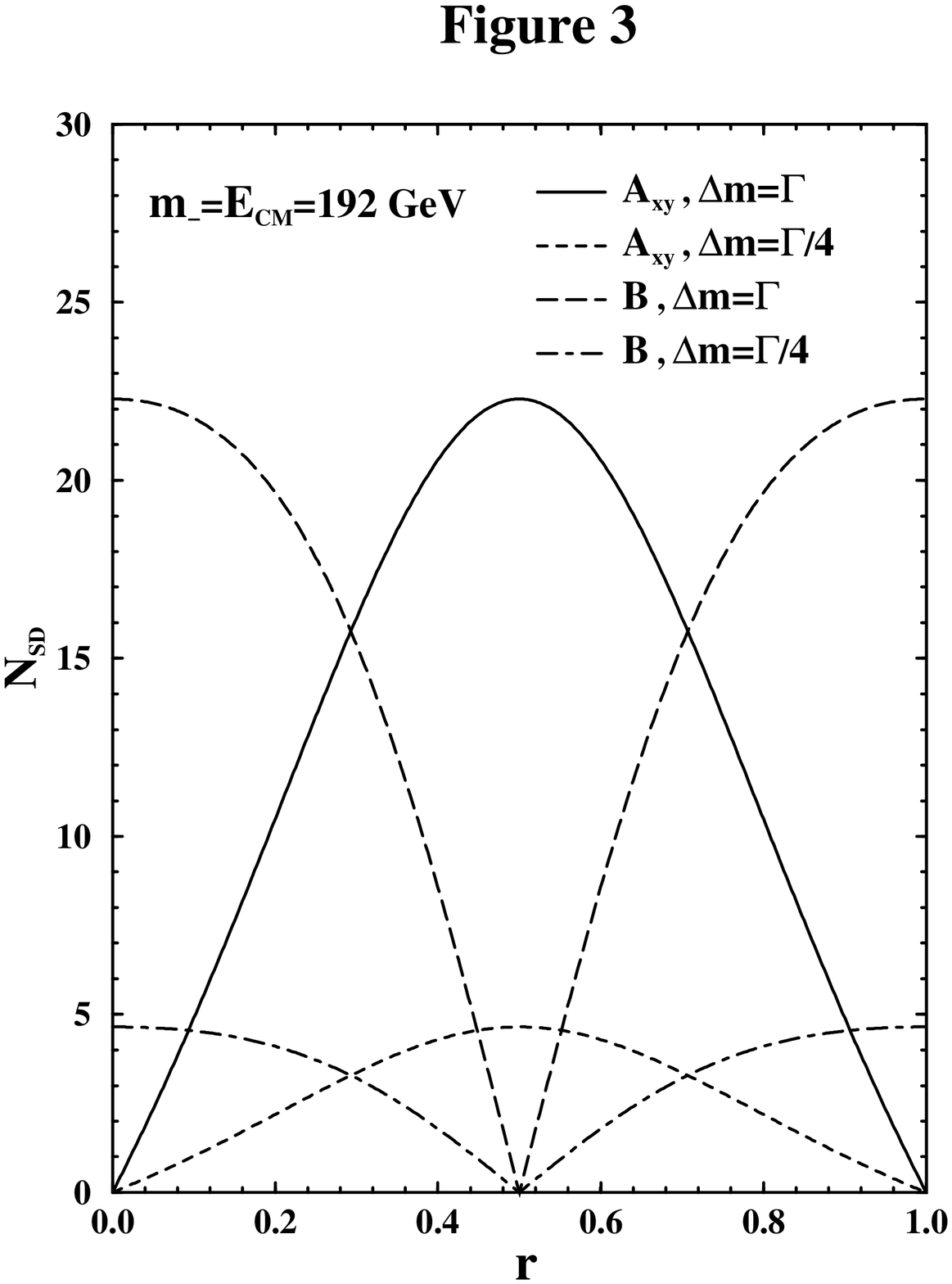}
\end{figure}


\begin{thebibliography}{99}
\bibitem{foot2} $T_N$ is 
defined as usual by replacing ${\rm time} \to - {\rm time}$ 
without changing initial and final states.
\end{thebibliography}
\end{document}